# Effects of Weak Links in the Nonlinear Microwave Response of MgB$_2$ Superconductor

A. Agliolo Gallitto[1], G. Bonsignore[1], G. Giunchi[2] and M. Li Vigni[1]

[1]*CNISM and Dipartimento di Scienze Fisiche ed Astronomiche, Università di Palermo, via Archirafi 36, 90123 Palermo, Italy*

[2]*EDISON SpA, R&D Division, Foro Buonaparte 31, 20121 Milano, Italy*

**Abstract**

We report experimental results of second-harmonic (SH) response at microwave frequency in several ceramic MgB$_2$ samples, prepared by different methods. The SH signal has been investigated as a function of the temperature and DC magnetic field. The investigation has been carried out at low magnetic fields, where nonlinear processes arising from motion of Abrikosov fluxons are ineffective. We show that the low-temperature SH emission is ascribable to processes involving weak links. Comparison among the peculiarities of the SH signal radiated by the different samples shows that the presence of weak links strongly depends on the sample preparation method, as well as the purity and morphology of the components used for the preparation.

**Key words**: Microwave response; Harmonic generation; Magnesium diboride; Weak links.

**Corresponding author**:

Prof. Aurelio AGLIOLO GALLITTO

Dipartimento di Scienze Fisiche ed Astronomiche
via Archirafi 36, I-90123 Palermo (Italy)
Tel: +39 091 6234207 – Fax: +39 091 6162461
Web: www.fisica.unipa.it – E-mail: agliolo@fisica.unipa.it

# 1. INTRODUCTION

Since the discovery of superconductivity at ~ 40 K in magnesium diboride, several authors have been devoting the attention to understand the origin of superconductivity in this material, as well as to explore its potential in technological applications [1-3]. Since the first studies, it has been shown that grain boundaries do not appreciably affect the transport properties of this material. In particular, in contrast to what occurs in cuprate high-$T_c$ superconductors, the critical current of bulk $MgB_2$ does not strongly depend on the magnetic field [1, 4]. Different authors have verified this property, very important for applications, and it has been suggested that grain boundaries in such superconductor do not act as weak links. It has been suggested that the different effect of grain boundaries in cuprates and $MgB_2$ may be related to the large coherence length of $MgB_2$, which would exceed the supposedly narrow grain-boundary width. However, some authors have observed that the amorphous regions of the grain boundaries, consisting of metallic materials, extend up to 500 Å wide, larger than the coherence length [5]. On the other hand, recent studies in bulk $MgB_2$ have highlighted voltage-flux oscillations due to the rf-SQUID effects, as well as harmonic generation at frequency of the order of 10 kHz, both related to the presence of weak links [6, 7]. However, the authors provide evidence that only a small amount of grain boundaries act as weak links.

Weak links in superconducting samples are undesirable for many reasons; they are responsible for the limitation of the critical current, give rise to large residual surface resistance at low temperatures, and are source of nonlinear effects in the electromagnetic response up to microwave (mw) frequencies [8-22]. Nonlinear mw effects have been highlighted from measurements of the surface resistance as a function of the input power [8-10], intermodulation product [11] and harmonic generation [12-23]. Several mechanisms responsible for the nonlinear response have been recognized, whose effectiveness depends on the temperature, magnetic field and type of superconductor [8-23]. In particular, it has been shown that at low magnetic fields and low temperatures the main source of nonlinearity arises from processes occurring in weak links [8, 13-18].

In this paper, we report results of second-harmonic (SH) emission at mw frequency by ceramic $MgB_2$ samples, prepared by different methods. The signal radiated at the SH frequency of the driving field has been investigated as a function of temperature and external magnetic field. In order to check the presence of weak links in the different samples, we have devoted the attention to the properties of the SH response at applied magnetic fields smaller than the lower critical field, $H_{c1}$, when weak links are not decoupled by the presence of Abrikosov fluxons. Although grain boundaries in $MgB_2$ do not appreciably affect the transport properties, we show that, even in this superconductor, the SH emission at low magnetic fields and low temperatures is due to processes occurring in weak links. Comparison among the peculiarities of the SH signal radiated by the different samples has shown that the presence of weak links strongly depends on the sample preparation method, as well as the purity and morphology of the components used for the preparation.

# 2. ESPERIMENTAL APPARATUS AND SAMPLES

## 2.1 Samples

The SH emission has been investigated in several samples of ceramic $MgB_2$, prepared by different methods. In table I we report the critical temperature, transition width and preparation method of the investigated samples. In the following, MgB$_2$ bulk samples are indicated as B; they have parallelepiped shape of comparable dimensions ($\approx 3 \times 2 \times 1$ mm$^3$).

Sample P$_\alpha$ consists of 5 mg of commercial Alfa-Aesar MgB$_2$ powder placed in a plexiglas holder. Sample B$_1$ has been obtained after sintering Alfa-Aesar powder at 800 °C in Ar atmosphere for 3 h. From AC susceptibility measurements at 100 kHz, it results that both P$_\alpha$ and B$_1$ have most likely an inhomogeneous superconducting transition, with two characteristic temperatures about 1.5 K apart [24].

Sample B$_2$ has been synthesized, at the Institute of Solid State Physics of the Russian Academy of Sciences (Chernogolovka), by direct reaction of B powder and a lump of Mg metal (both better than 99.95 % purity) at 1100 °C and subsequent rapid cooling [25]. It has been shown that MgB$_2$ samples prepared by this method contain some amount of Mg and MgO.

Sample B$_3$ has been prepared, at the INFM-LAMIA/CNR laboratory (Genova), using the so called one-step method; crystalline B (99.97 % purity) and Mg (99.999 % purity), put in Ta crucibles welded in Ar atmosphere and closed in quartz tubes under vacuum, were heated up to 950°C [26]. The sample undergoes a very narrow superconducting transition, indicating high sample homogeneity; X-ray diffraction pattern does not exhibit peaks due to the presence of extra phases, as Mg, MgO or MgB$_4$.

Samples B$_{4.1}$ and B$_{4.3}$ have been prepared by the reactive liquid Mg infiltration technology [27], which consists in the reaction of boron powder (commercial purity, Starck H. C., Germany) and pure liquid Mg. The reacting elements are sealed in a stainless steel container, with a weight ratio Mg/B over the stoichiometric value ($\approx$ 0.55). In the sealing operation, made by conventional tungsten inert gas welding procedure, some air remais trapped inside the boron powder. The two samples have been prepared using different boron powders: for sample B$_{4.1}$ it has been used microcrystalline B powder (98% purity, Am. Grade I) and the thermal annealing was done at 900°C for 30 min; for sample B$_{4.3}$ it has been used crystalline B powder (99.5% purity, original chunks mechanically grinded and sieved under a 100 micron sieve) and the thermal annealing was done at 900°C for 2 h. It has been already shown that the size of the used B powder greatly affects the morphology and the superconducting characteristics of the final material obtained by reactive liquid infiltration [28]. In particular, it has been shown that the final products consist in grains having the same dimension as the starting B powder embedded in a finer grained matrix. Fig.1 shows SEM micrographs of typical morphology of the MgB$_2$ materials, obtained by this method, using different B powders: (a) micron-size powder, as sample B$_{4.1}$; (b) crystalline B powder of size < 100 µm, as sample B$_{4.3}$. By microanalysis on the materials prepared as sample B$_{4.3}$, it has been shown that they have a composite structure consisting of three crystalline phases: MgB$_2$, Mg$_2$B$_{25}$, Mg. Among the impurity phases, the metallic Mg has been revealed in the inter-grain regions of the small grains and the Mg$_2$B$_{25}$ only in the intra-grain regions of the larger grains. On the contrary, materials prepared as sample B$_{4.1}$ consist on micron-size grains surrounded by narrow layers of metallic Mg of nanometric thickness [27, 28].

## 2.2 Experimental apparatus

The measurements of SH emission have been performed by a nonlinear spectrometer, operating at 3 – 6 GHz. The sample is placed in a bimodal cavity, resonating at the two frequencies ν and 2ν, with ν $\approx$ 3 GHz, in a region in which the mw magnetic fields *H*(ν) and *H*(2ν) are maximal and parallel to each other. The ν-mode of the cavity is fed by a train of mw pulses, with pulse width 5 µs, pulse repetition rate 200 Hz, maximum peak power

~ 50 W. The SH signal is detected by a superheterodyne receiver, with a sensitivity of ~ -75 dBm. The cavity is placed between the poles of an electromagnet that generates DC magnetic field, $H_0$, up to $\approx$ 10 kOe; two additional coils, externally fed, allow reducing the residual field within 0.1 Oe and working at low magnetic fields. All the measurements have been performed with $\boldsymbol{H}_0 \parallel \boldsymbol{H}(\nu) \parallel \boldsymbol{H}(2\nu)$. Further details of the experimental apparatus are reported in Ref.[29].

## 3. EXPERIMENTAL RESULTS

The SH signal has been investigated as a function of the temperature and the DC magnetic field. In order to recognize possible SH signals related to nonlinear processes in weak links, the attention has been devoted to the SH response at low external fields ($H_0 < H_{c1}$).

Fig. 2 shows the SH signal intensity as a function of the temperature in the $MgB_2$ samples, at $H_0 = 10$ Oe. Before the measurement was performed, the sample was zero-field cooled (ZFC) down to $T = 4.2$ K, then $H_0$ was set at 10 Oe and kept constant during the measurement. As one can see from plot (a), in both $P_\alpha$ and $B_1$ the SH emission is significant in the whole range of temperatures investigated and exhibits a peak below, but very near to, $T_c$. The temperature dependence of SH signal is similar for the two samples, except for the peak intensity, which is higher in the powder. The two curves show a kink at $T \approx 38$ K, which reflects the inhomogeneity of the superconducting transition.

Comparison among the results obtained in the different samples shows that in $B_2$, $B_3$ and $B_4$ samples the SH signal is weaker than those detected in the samples prepared by Alfa-Aesar powder (please note the different scale of plot (a)). Furthermore, the intensity of the near-$T_c$ peak is roughly the same in $B_2$, $B_3$ $B_{4.1}$ and $B_{4.3}$; on the contrary, the low-$T$-signal intensity strongly depends on the specific sample, resulting undetectable in $B_{4.1}$. It is worth noting that, although the samples exhibit superconducting transitions of very different widths, the near-$T_c$ peaks have roughly the same width in all the samples except in $B_{4.1}$.

The peculiarities of the SH signal at temperatures near $T_c$ have been studied as a function of $H_0$ and the input power; they are very similar in all the samples investigated. The results can be summarized as follows: i) on increasing the DC magnetic field, the peak broadens and its maximum shifts toward lower temperatures, ii) the peak width does no appreciably depend on the input power level, iii) the power dependence of the SH signal intensity is less than quadratic. A detailed study for sample $B_{4.1}$ is reported in Ref. [30].

Fig. 3 shows the field dependence of the SH signal at $T = 4.2$ K, observed in the ZFC $B_{4.3}$ sample by cycling the magnetic field in the range $-H_{max} \div +H_{max}$. The signal shows a hysteretic loop having a butterfly-like shape; on increasing the value of $H_{max}$, the hysteresis is more and more enhanced and the minima, observed at low fields, move away from each other. The hysteretic loop maintains the same shape in subsequent field runs as long as the value of $H_{max}$ is not changed. It is worth remarking that no hysteresis is observed when $H_{max}$ is smaller than about 2 Oe; furthermore, the low-field structure disappears irreversibly when $H_{max}$ reaches values of the order of 100 Oe.

Results similar to those of Fig. 3 have been obtained in the other samples. Fig. 4 shows the SH signal intensity as a function of $H_0$ in the different samples, obtained at $T = 4.2$ K by cycling the magnetic field between $\approx \pm 20$ Oe. The curve for $B_1$ sample has not been reported because it is very similar to that obtained in $P_\alpha$. The curves show qualitatively the same behavior, even if they differ in some peculiarities, such as the intensity and the position of the maxima and minima. In all the samples, this low-field structure disappears irreversibly after that the sample is exposed to a magnetic field of the order of the lower critical field; this finding strongly suggests that the process responsible for the low-$T$ and low-field SH signal becomes ineffective when Abrikosov fluxons penetrate in the samples.

## 4. DISCUSSION

The nonlinear response of high-$T_c$ superconductors to electromagnetic fields has been extensively investigated [8-23, 31-34]. At low temperatures, it has been ascribed to extrinsic properties of the superconductors such as impurities, weak links [8, 13-18, 31] or fluxon motion [8, 14-17, 34]. On the contrary, nonlinearity at temperatures close to $T_c$ can be related to intrinsic properties of the superconducting state; in particular, it has been ascribed to modulation of the order parameter induced by the mw field [14, 15, 19, 20] or, alternatively, to nonlinear Meissner effects [11, 21-23]. Investigation of mw harmonic emission performed up to now in ceramic $MgB_2$ samples strongly suggested that, also in this class of superconductors, different mechanisms come into play in the mw harmonic emission [24, 29, 30, 35].

A detailed study of second- and third-harmonic emission in Alfa-Aesar powder [24, 35] has shown that the low-$T$ and near-$T_c$ harmonic signals have different origin. In particular, investigation of the peculiarities of the SH signal at different input power levels [35] has confirmed that the near-$T_c$ peak cannot be ascribed to processes occurring in weak links. On the contrary, by properly handling the Alfa-Aesar powder, we have shown that the SH signal observed at low fields and low temperatures is ascribable to the presence of weak links [35].

As one can see from Fig. 4, the low-$T$ SH signal exhibits qualitatively the same properties in all the samples; so, we infer that weak links may be present also in ceramic $MgB_2$ samples. In the presence of weak links, two nonlinear processes may come into play. Harmonic emission is expected when supercurrents are induced by the DC and mw magnetic fields in loops containing Josephson junctions. In this case, harmonic emission is strictly related to the intrinsic nonlinearity of the Josephson current [16, 31]. In addition, inter-grain dynamics of Josephson fluxons in the critical state may give rise to harmonic emission [17, 32, 33]. On the other hand, the hysteretic behavior of the low-$T$ SH signal suggests that processes due to trapped flux are involved in the SH emission. Since we observe magnetic hysteresis at fields of the order that 10 Oe, much lower than $H_{c1}(T)$, we infer that a critical state of Josephon fluxons develops in the samples. Results similar to those reported in Figs. 3 and 4 have been obtained in $Ba_{0.6}K_{0.4}BiO_3$ crystals [17, 18]; they have been justified supposing that both the above-mentioned nonlinear processes involving weak links come into play simultaneously. The hypothesis that the low-$T$ and low-field SH signal is ascribable to processes occurring in weak links is further supported by the irreversible loss of the low-field structure, observed in ZFC samples, after the sample is exposed to high fields. Indeed, when $H_0$ reaches values higher than $H_{c1}$, Abrikosov fluxons penetrate the grains and the Josephon junctions are decoupled by the applied field and/or the trapped flux.

The results of Fig. 2 show that in all the samples an enhanced SH emission is observed at temperatures near $T_c$. Although the investigated samples exhibit superconducting transitions of different width, the peak in the SH-$vs$-$T$ curves has roughly the same width for $P_\alpha$, $B_2$, $B_3$ and $B_{4.3}$ samples; the peak is narrower in $B_{4.1}$ and wider in $B_1$. We think that the mix of the near-$T_c$ signal with that arising from weak links may affect the peak width; so, the peak is narrower in $B_{4.1}$ because this sample does not exhibit any SH signal due to weak links, the opposite occurs in $B_1$. In order to recognize the origin of the SH emission at temperatures close to $T_c$, we have performed a comprehensive investigation in sample $B_{4.1}$. The results have been discussed in the framework of a phenomenological model, which assumes that at temperatures close to $T_c$ the mw and the DC fields, penetrating in the surface layers of the sample, perturb the partial concentrations of the normal and superconducting electrons [30].

The presence of SH emission due to nonlinear processes in weak links, highlighted by our results, seems to be in conflict with the results reported in the literature by several authors, who show that grain boundaries in $MgB_2$ do not worsen the transport properties of the materials [1, 4]. On the other hand, recent studies in bulk $MgB_2$ have highlighted voltage-flux oscillations due to the rf-SQUID effects, as well as harmonic generation at 10 kHz, related to the presence of weak links [6, 7]. However, the authors provide evidence that only a small amount of grain boundaries act as weak links, contrary to what occurs in cuprate superconductors. As one can see from Fig. 2, in the most of the bulk samples, here investigated, the low-$T$ SH signal is less intense than in the powder; this indicates that only few of the natural grain boundaries are effective in the nonlinear response (a sample of ceramic YBCO of comparable dimension would exhibit a SH signal several order of magnitude greater [18, 29]). In sample $P_\alpha$, Josephson junctions from the contacts between the powder grains may contribute to the nonlinear emission, giving rise to an intense SH signal. In sample $B_1$, which has been obtained by Alfa-Aesar powder, the SH-signal intensity is of the same order as $P_\alpha$; this suggest that the sintering process did not strongly connect the powder grains. Actually, it is not well understood the reason way $B_3$ sample exhibits harmonic signal from weak links; indeed, as it has been discussed in Section 2.1, the sample has been prepared using very pure elements by the one-step method, which generally gives rise to $MgB_2$ samples having well-connected grains [25]. However, we would note that the low-$T$ SH signal in this sample is very weak.

Concerning the different SH response of samples $B_{4.1}$ and $B_{4.3}$, both produced by reactive Mg liquid infiltration, it can be accounted for considering the different purity and morphology of the two samples (see Fig.1 and Section 2.1). Weak links in $B_{4.3}$ may be present in both the fine-grained matrix, due to the presence of thick metallic Mg, as well as inside the larger grains, due to the presence of the two crystalline phases: $MgB_2$, $Mg_2B_{25}$. On the contrary, $B_{4.1}$ is a highly homogeneous sample constituted by micron-size grains of $MgB_2$ surrounded by narrow layers of metallic Mg of nanometric thickness, which do not act as weak links.

From these considerations, one can qualitatively justify the different intensities of the SH signal due to weak links, as well as the different shape of the hysteretic loops of Fig. 4, observed sweeping the DC magnetic field. Indeed, the hysteretic behavior is most likely related to critical state of Josephson fluxons in the weak links and, therefore, it strongly depends on the morphology of the weak links, which can be different in the different samples.

## 4. CONCLUSION

We have reported a set of experimental results on second-harmonic generation in several ceramic $MgB_2$ samples, prepared by different methods. The signal radiated at the second-harmonic frequency of the driving field has been investigated as a function of temperature and DC magnetic field. The attention has been devoted to the response at applied magnetic fields smaller than the lower critical field, when nonlinear processes related to motion of Abrikosov vortex do not play a role in the harmonic emission. In most of the investigated samples, the signal intensity is roughly constant from low temperatures up to few K below $T_c$ and shows an enhanced peak near $T_c$. The low-$T$ signal originates from processes occurring in weak links, while the enhanced emission near-$T_c$ is due to modulation of the normal and superconducting fluid densities, induced by the mw field. A comparison among the results obtained in the different samples has shown that the intensity of the low-$T$ signal strongly depends on the preparation method as well as the purity and morphology of the components used for the preparation.

In conclusion, although grain boundaries in $MgB_2$ do not appreciably affect the transport properties of the superconducting material, even in this superconductor the second-harmonic emission at low magnetic fields and low temperatures is due to processes occurring in weak links. The reason for the different behavior of the natural grain boundaries of $MgB_2$ in DC and AC response is not yet fully understood. We suggest that the harmonic-generation technique allows highlighting the presence of weak links because it selects only the response of the grain boundaries that are weakly linked and it is not affected by the linear response of the grain boundaries strongly coupled. We would remark that in bulk samples of $MgB_2$ prepared using pure components, the harmonic emission is noticeable reduced with respect to that observed in ceramic cuprate superconductors. This property is of relevance for using $MgB_2$ in the passive mw devices.

**ACKNOWLEDGMENTS**

The authors are very glad to thank N. N. Kolesnikov, M. P. Kulakov and P. Manfrinetti for having kindly supplied some $MgB_2$ samples and G. Lapis and G. Napoli for technical assistance.

**TABLE CAPTIONS**

Table I: Critical temperature, transition width (90 % – 10 % criterium) and preparation method of the investigated sample. Bulk samples are indicated by B.

**FIGURE CAPTIONS**

**Fig. 1**: SEM micrographs of typical morphology of the $MgB_2$ materials, obtained by reactive liquid Mg infiltration method using different B powders: (a) micron-size powder, as sample $B_{4.1}$; (b) crystalline B powder of size < 100 µm, as sample $B_{4.3}$.

**Fig. 2**: SH signal intensity as a function of the temperature for the $MgB_2$ samples. DC magnetic field $H_0$ = 10 Oe; input peak power $\approx$ 30 dBm. Please note the different scale of plot (a). The noise level corresponds to $\approx$ –75 dBm.

**Fig. 3**: Magnetic-field dependence of the SH signal, observed in the ZFC $B_{4.3}$ sample by cycling the magnetic field in the range $- H_{max} \div + H_{max}$. $T$ = 4.2 K; input peak power $\approx$ 30 dBm. The three curves refer to different values of $H_{max}$. Lines are leads for eyes.

**Fig. 4**: SH signal intensity as a function of $H_0$ in the different samples, obtained by cycling the magnetic field between $\approx \pm 20$ Oe. $T$ = 4.2 K; input peak power $\approx$ 30 dBm. Lines are leads for eyes.

| Sample | Onset $T_c$ (K) | $\Delta T_c$ (K) | Preparation method |
|---|---|---|---|
| P$_\alpha$ | 38.5 | 7.2 | Commercial Alfa-Aesar powder |
| B$_1$ | 38.5 | 6.4 | Sintering Alfa-Aesar powder at 800 °C in Ar atmosphere for 3 h |
| B$_2$ | 38.4 | 1.3 | Direct reaction of B powder and a lump of Mg metal [25] |
| B$_3$ | 38.9 | 0.3 | One step method [26] |
| B$_{4.1}$ | 38.7 | 0.7 | Reactive liquid Mg infiltration technology [27] |
| B$_{4.3}$ | 39.0 | 1.1 | |

Table I

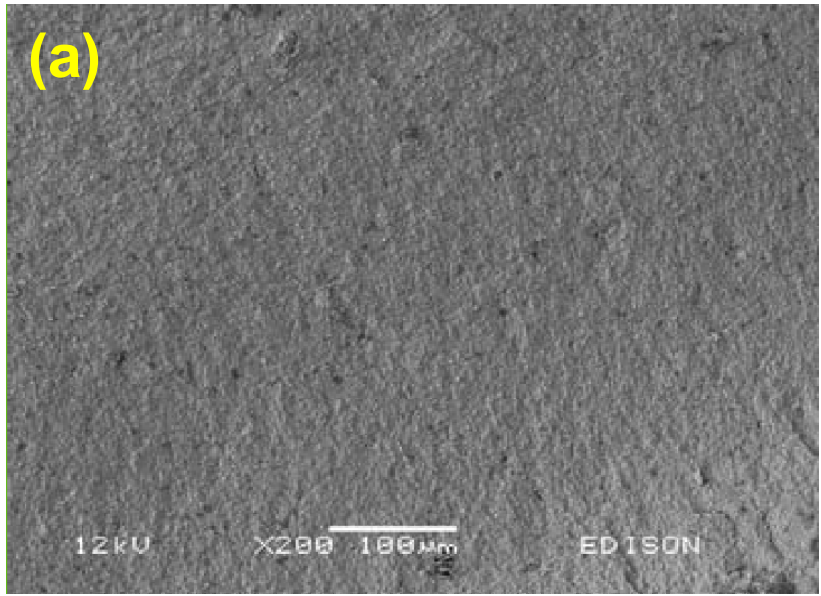

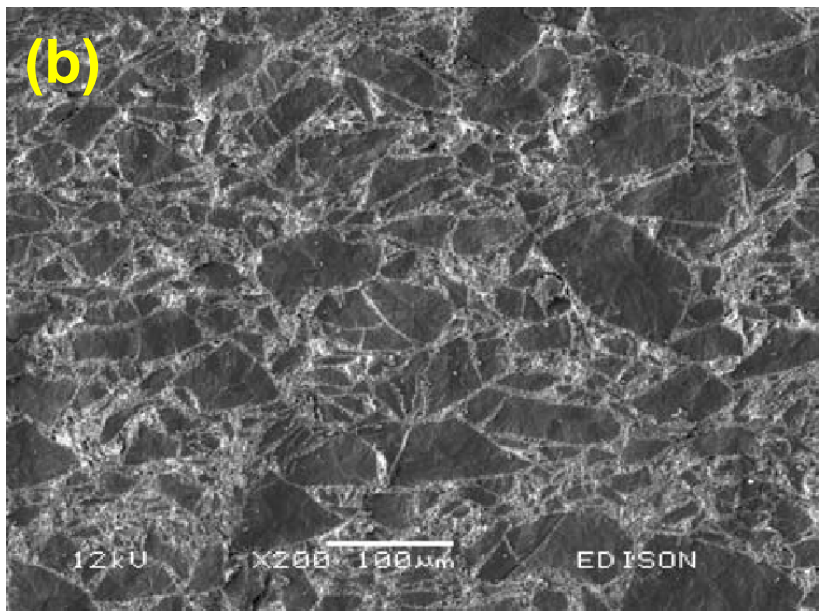

Figure 1

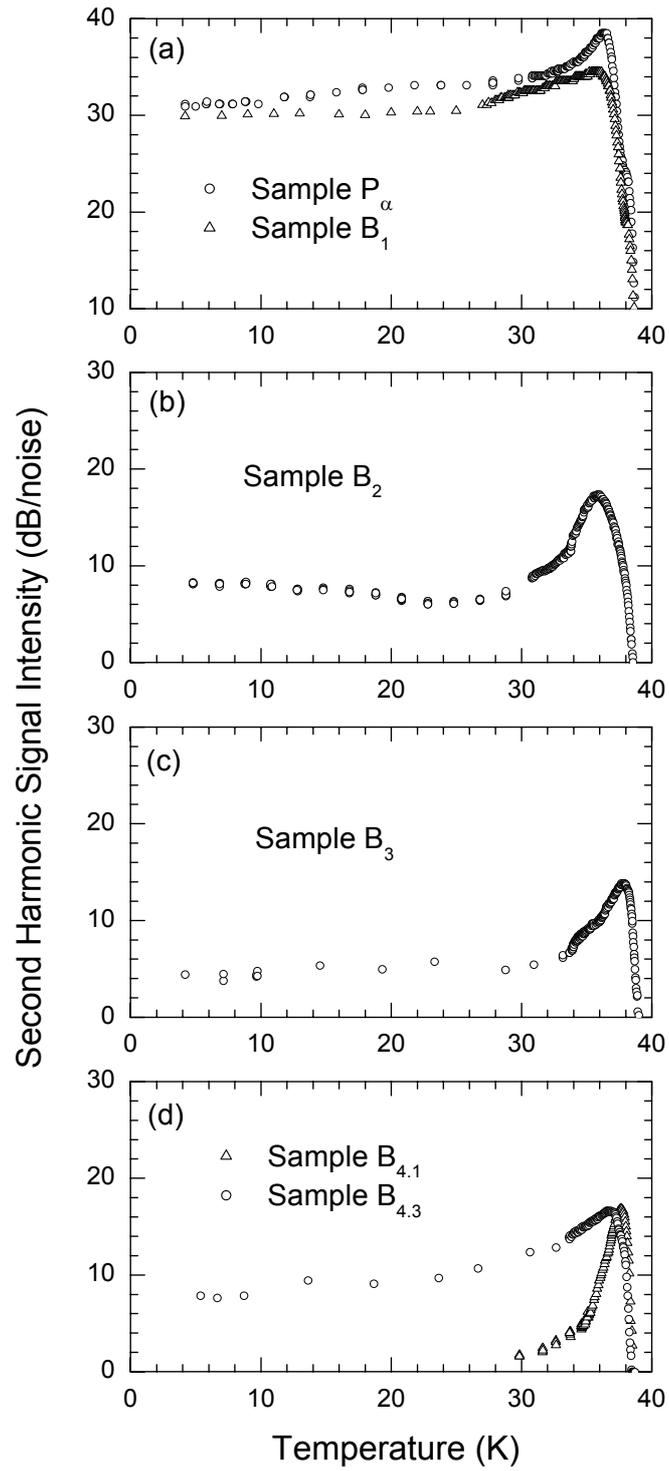

Figure 2

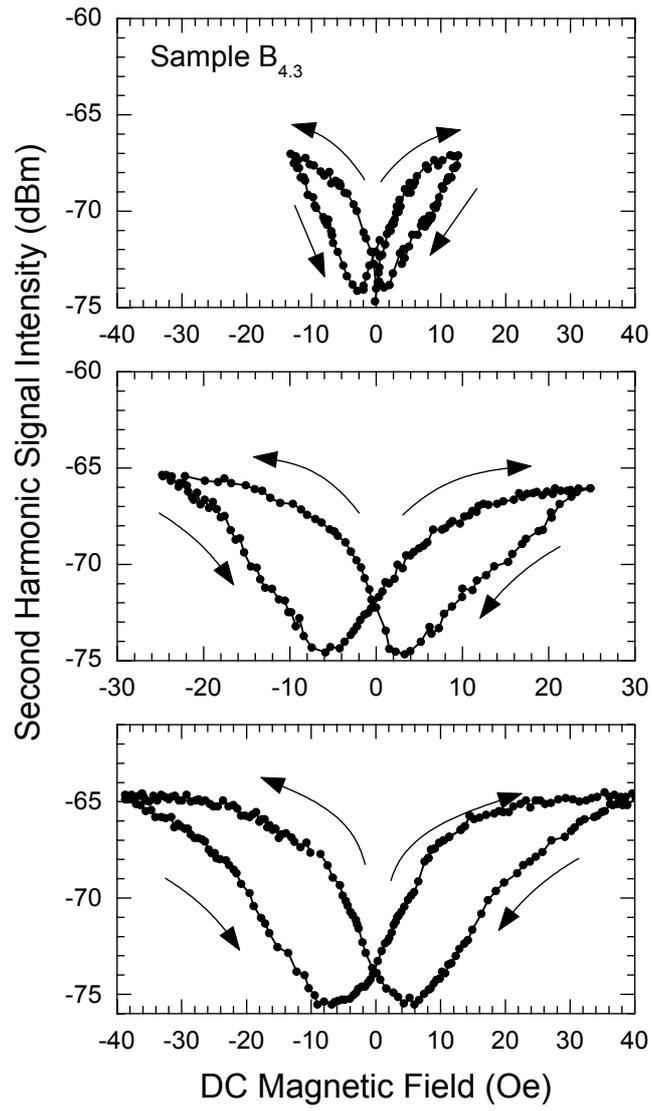

Figure 3

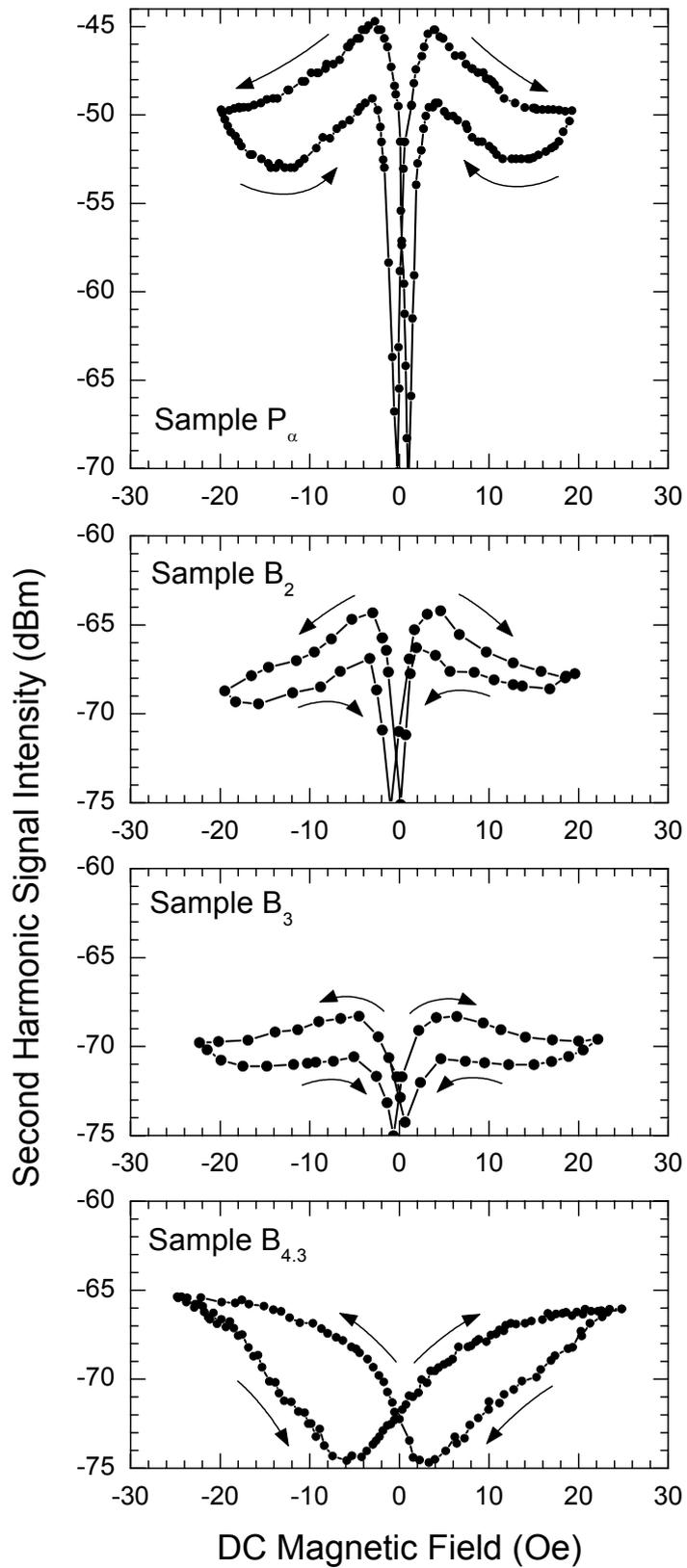

Figure 4